# A Novel Method of Encoded Multiplexing Readout for Micro-pattern Gas Detectors [1]

QI Bin-Xiang[1,2]   LIU Shu-Bin[1,2]   JI Heng[1,2]   SHEN Zhong-Tao[1,2]   MA Si-Yuan[1,2]
LIU Hong-Bang[3]   HUANG Wen-Qian[3]   AN Qi[1,2]

[1] State Key Laboratory of Particle Detection and Electronics, University of Science and Technology of China, Hefei 230026, China

[2] Department of Modern Physics, University of Science and Technology of China, Hefei 230026, China

[3] University of Chinese Academy of Sciences, Beijing 100049, China

**Abstract**: The requirement of a large number of electronic channels poses a big challenge for Micro-pattern Gas Detector (MPGD) to achieve good spatial resolution. By using the redundancy that at least two neighboring strips record the signal of a particle, a novel method of encoded multiplexing readout for MPGDs is presented in this paper. The method offers a feasible and easily-extensible way of encoding and decoding, and can significantly reduce the number of readout channels. A verification test was carried out on a $5\times5$ cm$^2$ Thick Gas Electron Multiplier (THGEM) detector using a 8 keV Cu X-ray source with 100um slit, where 166 strips are read out by 21 encoded readout channels. The test results show a good linearity in its position response, and the spatial resolution root-mean-square (RMS) of the test system is about 260 μm. This method has an attractive potential to build large area detectors and can be easily adapted to other detectors like MPGDs.

## 1. Introduction

Advances in microelectronics and printed-circuit board (PCB) techniques during the past decades triggered a major transition in the field of gas detector from wire chambers to MPGDs. Due to the good spatial resolution, high rate capability, large active areas, and radiation hardness, MPGDs such as the Gas Electron Multiplier (GEM) [1], the Thick GEM(THGEM) [2] and the Micromegas [3] are widely used in high-energy and particle physics, they also opened a new trend in fundamental science, medical imaging and industry [4]. MPGDs need high-density narrow anode readout elements to achieve good spatial resolution. Most of the readout techniques employ a large number of electronic channels to readout directly. The large number of readout channels has become an issue for most experiments using MPGDs. Consequently, some alternative readout techniques have been employed to reduce the number of electronic channels, such as resistive interpolating readout [5][6], and delay-line readout [7][8].

Received 8 Sep. 2015

* Supported by National Natural Science Foundation of China (Grant No.11222552 and NO.11265003)

1) E-mail: qibx@mail.ustc.edu.cn

2) E-mail: liushb@ustc.edu.cn (corresponding author)

By using the redundancy that each particle usually showers the signal on several neighboring strips in MPGDs, an encoded multiplexing readout technique was developed by S.Procureur group at Saclay, France [9], which innovatively reduces the number of readout channels, but the encoding method is complicated and the encoding sequences need to be reordered. A novel method of encoded multiplexing readout is presented in this paper, which offers a simple and easily-extensible way of encoding, and it is feasible to decode the hit position where a signal is shared on k neighboring strips, k>=2. This method can dramatically reduce the number of readout channels, and it has been successfully tested with a 5×5 cm² THGEM [10] where 166 strips are read out by 21 encoded readout channels.

## 2. Principle and Method

### 2.1 Principle

The development of large areas and high spatial resolution of MPGDs requires a large number of readouts. However, for a particular event, the signal is usually localized on a few related electronic channels, the others being useless, especially in the low incident flux experiments. This feature of signal sparsity and the electronic channels redundancy can be used to track the particles with an appropriate encoded multiplexing pattern, which could reduce a large number of electronics.

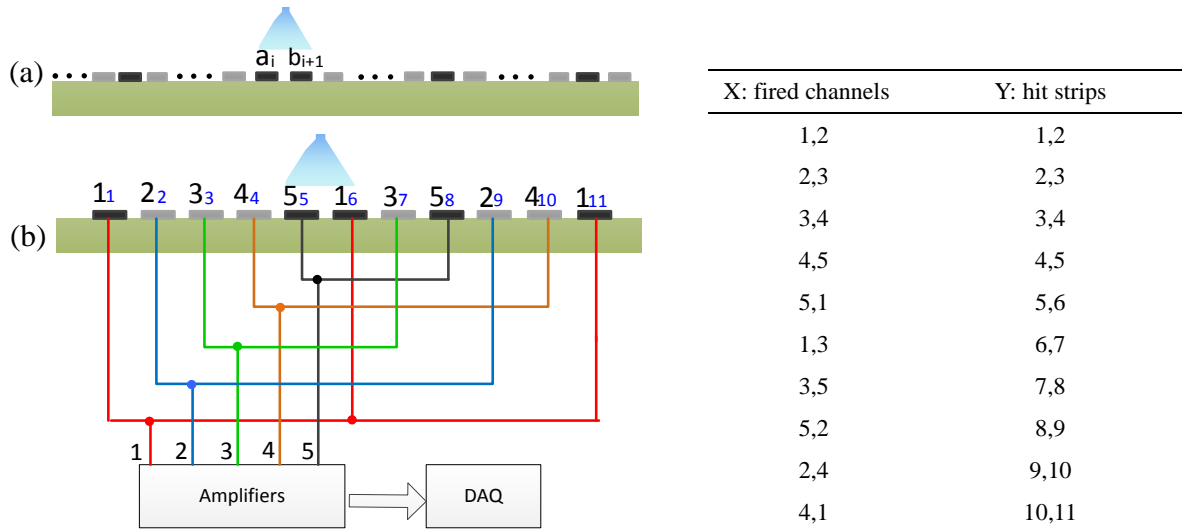

Fig. 1. Principle of the encoded multiplexing method.    Table 1 Decoding table of 5 readout channels

| X: fired channels | Y: hit strips |
| --- | --- |
| 1,2 | 1,2 |
| 2,3 | 2,3 |
| 3,4 | 3,4 |
| 4,5 | 4,5 |
| 5,1 | 5,6 |
| 1,3 | 6,7 |
| 3,5 | 7,8 |
| 5,2 | 8,9 |
| 2,4 | 9,10 |
| 4,1 | 10,11 |

Because of the charge transverse diffusion, a particle signal almost showers on at least two neighboring strips in MPGDs. This feature can be utilized to track the particles with an appropriate encoded multiplexing construction {strips}→{channels}. As shown in Fig. 1(a), where $X_Y$ represents the each encoded multiplexing connection, X is the channel number and Y is the strip number. All the channel numbers X could form an encoding list by the sequence of the connected strip numbers (Y). Supposed that two neighboring strips i and i+1 are connected to two given channels a and b, and the two channels are also connected to several other non-neighboring strips. If a signal is recorded only on channels a and b, it is almost certain that the hit position is in the strips i and i+1, because of there is only one doublet combination of a and b in the encoding list. In other words, if any doublet combination of channel numbers appears at most once in the encoding list, the hit positon of signal can be precisely tracked by the case of two neighboring fired strips.

A specific example is shown in Fig. 1 (b), where 11 strips are read out by 5 readout channels. All the $C_5^2$ doublets combinations of 5 channels appears once in the encoding list{1,2,3,4,5,1,3,5,2,4,1}, which corresponds to the 11strips. A particle event hits in two neighboring strips 5 and 6, which results the signal is recorded on its encoded channel 5 and 1. In turn, the combination of fired channel 5 and 1 can uniquely decode the hit position strip 5 and 6. The combination of two fired channels can uniquely decode the hit strips of the particle in the detectors, as shown in Table 1.

Similarly, if all the $C_n^2$ doublets combinations of n channels are constructed to a feasible encoding list in an appropriate way, $C_n^2$ doublets two neighboring strips can be uniquely decoded. It means that n channels can read out theoretical maximum of $C_n^2+1$ strips, as $C_n^2+1$ strips contain $C_n^2$ doublets two neighboring strips.

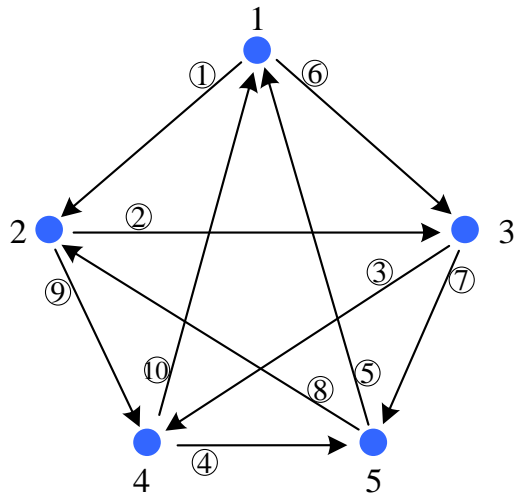

Fig. 2.    An Euler Walk of 5 readout channels.

Generally, the principle described above is a graph theory problem that whether there is an Euler walk to construct a path which visits each edge exactly once, where the doublet combinations represent the edges and the readout channels represent the vertices. According to Euler's path theorem [11], it can be proved that there is an Euler walk when the numbers of readout channels n is an odd number, as all of its vertices have even degree. In other words, n channels can construct a maximum encoding list of $C_n^2+1$ strips when n is an odd number, and the each doublet combination occurs exactly once in the list. Fig.2 shows an Euler Walk of 5 readout channels, corresponding with the encoding list {1,2,3,4,5,1,3,5,2,4,1} of Fig.1 (b). It turns out that there are more than one constructions of Euler walk, such as 5 channels have the other encoding lists {1, 2, 3, 1, 4, 2, 5, 3, 4, 5, 1} and {1, 3, 5, 2, 4, 1, 2, 3, 4, 5, 1}.

## 2.2    Encoding

A practical case must be considered that a detector records a signal on more than two neighboring strips while the previous discussion based on the assumption of only two neighboring fired strips. It triggers the question of k-uplets (k>2) repetition which may lead to incorrectly decoding. As is seen in Fig.1, when a signal is recorded on the channel 2,3,4,5, it is not sure whether the hit position in $\{2_2,3_3,4_4,5_5\}$ or $\{3_7,5_8,2_9,4_{10}\}$. An alternative solution is to choose an optimized encoded multiplexing connection with appropriate constraints so as to minimize the deviation of hit position.

To obtain an available encoding list for the general case that a signal is recorded on k neighboring strips, k>=2，encoding constraints and rules are made as follows：

- use an odd number of electronics channel to encoding readout.
- any doublet combination of channel numbers appears exactly once in the whole list, and they are constructed as an Euler walk.
- the encoded connections are listed by row, and add one row on the list when two new channel are added.
- for the k-row, the k-list is the new added connections by channel 2k and 2k+1,where 2k an 2k+1 are interleaved in{1,2,3….2k-1}then form the k-list{1,2k,2,2k+1,3,2k,… 2k-1,2k,2k+1}.

The encoding list of multiplexing connections constructed by the above rules is shown in Table 2. The form $X_Y$ represents each multiplexing connection, where X is the channel number and Y is the strip number along the detector. For 2k+1 readout channels, the encoding list consists of k rows, corresponding with $C_{2k+1}^2 + 1$ strips as shown in Fig.3.

Table. 2.   The encoding list of multiplexing connections for 2k+1 readout channels

| row | the list of encoded multiplexing connections |
|---|---|
| 1 | $1_1, 2_2, 3_3$ |
| 2 | $1_4, 4_5, 2_6, 5_7, 3_8, 4_9, 5_{10}$ |
| 3 | $1_{11}, 6_{12}, 2_{13}, 7_{14}, 3_{15}, 6_{16}, 4_{17}, 7_{18}, 5_{19}, 6_{20}, 7_{21}$ |
| . | ................................. |
| k | $1_{C_{2k-1}^2+1}, (2k)_{C_{2k-1}^2+2}, 2_{C_{2k-1}^2+3}, (2k+1)_{C_{2k-1}^2+4}, 3_{C_{2k-1}^2+5}, (2k)_{C_{2k-1}^2+6}…..(2k-1)_{C_{2k+1}^2-2}, (2k)_{C_{2k+1}^2-1}, (2k+1)_{C_{2k+1}^2}$ $1_{C_{2k+1}^2+1}$ |

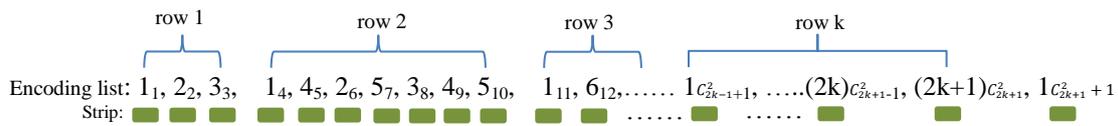

Fig. 3   The corresponding relation between the encoding list and strips

## 2.3   Decoding

In the above construction, the designed encoding list is robust to decoding the hit position of two fired channels. For a given doublet of two fired readout channels (a,b), the larger number can fix the row position so that the exact fired strip can be decoded with another number. For example, the doublet of channels (7,5) can be uniquely decoded to the strips (18,19).

What is important is to verify the feasibility of decoding the general case of k neighboring fired strips, k>2. By using the encoding rules and encoding list in Table.2, it can be analyzed as follows: when k=3, most of the position can be uniquely decoded except the end of each row. For example, the fired channel 5,6,7 is decoded to the hit position{$7_{18}$, $5_{19}$, $6_{20}$, $7_{21}$}, which results in 1 strip uncertainty of hit range. But it is explicit that there are no repetition of 3-uplets of 5,6,7 at other position. Similarly, it turns out there will be less than 2 strips uncertainty of hit range caused by the multiplexing decoding when k neighboring fired strips, k>2, but this will not lead to an incorrect decoding.

For the case where the signal records on k channels, k>2, it can be decoded by reappearing the sequence of the fired channels based on the encoding rules. For instance, the fired channel 1,2,3,6,7 is decoded to the hit range {$1_{11}$, $6_{12}$, $2_{13}$, $7_{14}$, $3_{15}$, $6_{16}$, } with 1 strip uncertainty. In addition, the uncertainty can be minished by the center method and center of gravity method.

Besides, the front rows of encoding list are constructed with the small numbers of readout channels because the list is added with every two new channels, which will leads to a large uncertainty. For instance, the fired channel 1,2,3,4,5,6,7, will be decoded to the hit range from strip 1 to strip 21 with 10 strips uncertainty .So the first few rows can be discarded to decode a larger k neighboring fired strips. For example, when discard the first three rows and the encoding list starts from the fourth row, this list can decode precisely at the case of at most 16 neighboring fired strips.

## 3  Verification Test with 5×5 cm² THGEM

### 3.1  Design of anode readout PCB

In order to verify this method, an encoded multiplexing anode readout PCB is manufactured and equipped for a THGEM detector with 5×5 cm². The anode readout PCB has 166 one-dimensional strips of 152 μm width and 304 μm pitch. As discussed previously, the first four rows of the encoding list are discarded in this design so as to decode at the case of at most 20 neighboring fired strips. Theoretically, it should be more than enough to the almost all signals of the detector. Therefore, the 166 strips need 21 channels for encoded multiplexing readout. The encoding list starts from (1,10…) of the fifth row and it is end up to (…20,16) of the tenth row. Each strip is connected to the corresponding channel based on the encoding list. Fig. 4 shows the corresponding routing between the strips and the readout channels on the PCB layout.

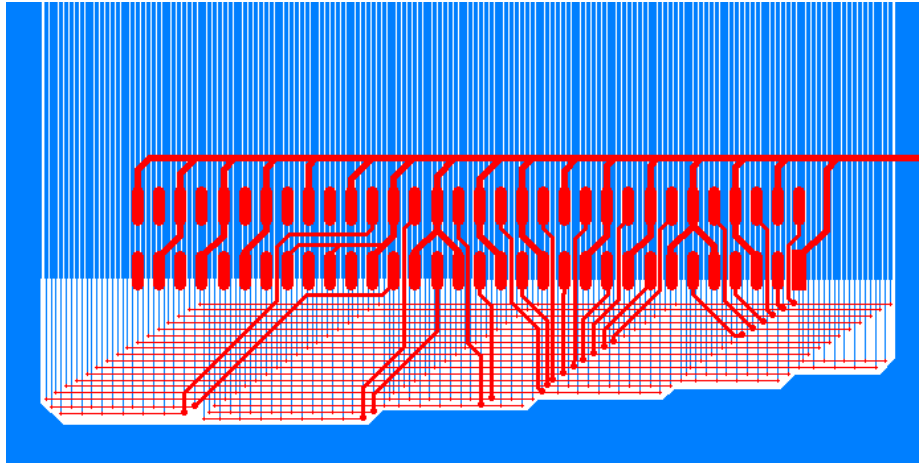

Fig. 4.   Scheme of the PCB connections between the 166 strips and 21 channels.

## 3.2  Experimental setup and results

As shown in Fig. 5, a verification test was carried out on the THGEM detector using a 8 keV Cu X-ray source and Ar/iso-butane (97:3) gas mixture. The detector was biased to a total gain of $1\times10^4$. A slit about 100 μm width in a thin brass sheet was used to produce a miniaturized X-ray beam. A manual movable platform was used for the postion scanning test.

Signals from 21 multiplexed channels were digitized by a GASTONE chip [12], then readout by a Xilinx development board to process and analyze. GASTONE is a 64-channel digital-output ASIC designed to readout the GEM Inner Tracker detector of K Long Experiment (KLOE). Each channel is made of a charge sensitive preamplifier, a shaper, a discriminator and a monostable module. Digital output data are transmitted via serial interface at 100 Mbit/s data rate, so the electronics can response a high event rate. With proper grounding and shielding, the threshold of GASTONE chip was set down to 100 mv in this test, corresponding to 5 fc.

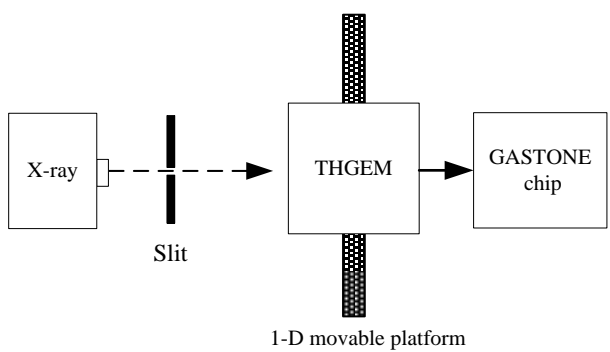

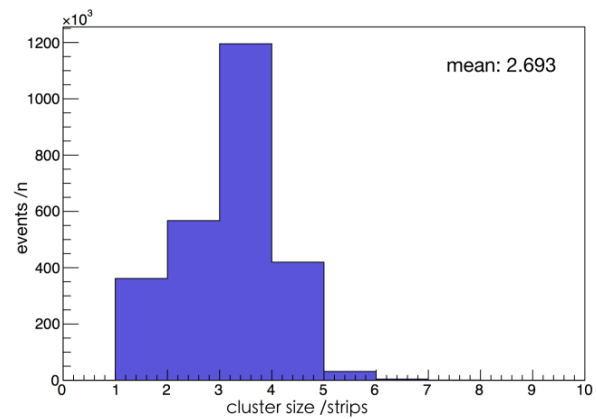

Fig. 5.   Experimental setup of the verification test.    Fig. 6.   Cluster size distribution of the signals.

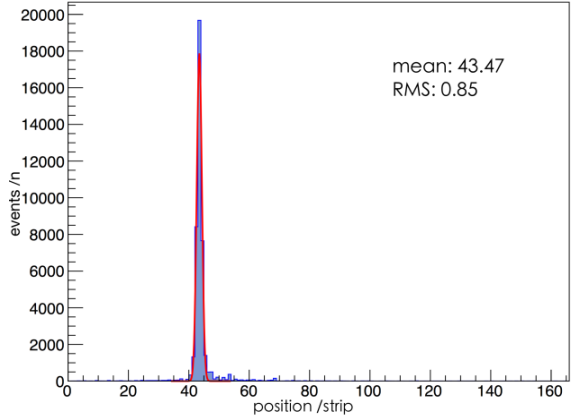 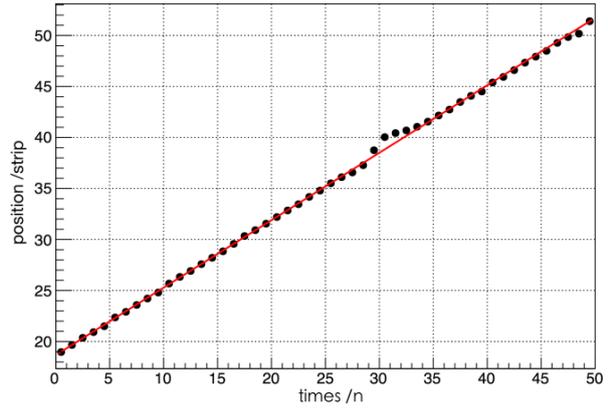

Fig. 7.  Position resolution measured for the upper part of detector.    Fig. 8.  Decoded results of postion scanning test.

A cluster size distribution before decoding procedure was tested and shown in Fig. 6. The mean cluster size is 2.69, and more than 86% of events can be tracked precisely on the detector, with a cluster size of greater than 1. Moderating the drift electric and improving S/N ratios can spread the cluster size, thus more events could be tracked.

The GASTONE is a digital output chip and can't be used to measure the charge value, so the center method is implemented in decoding procedure. Fig. 7 shows the decoded result that the signal hit in the position of mean strip 43.47 with RMS of 0.85 strip, matching with the fact that X-ray beam spot over the upper part of detector. During the position scanning test, the detector was moved at a step of 0.2mm in the 10 mm range. Fig. 8 shows the summary of the decoded position for the X-ray scan across the readout strip. There are some way-off point near the strip 40, this is because the encoding list of strips (39,40,41,42) is channels(13,11,12,13). As discussed above, it will result in at most 1 strip uncertainty. The uncertainty can be corrected in the analog-output front-end electronics by finding the largest set of consecutive fired channels in the encoding list.

The test results indicate that the method can correctly decode the hit position, and it has a good linearity in the postion scanning test.The spatial resolution RMS of the test system is about 260 μm (0.85 pitch),including the contributions of x-ray source,the detector and the noise.

## 4  Discussion and Conclusion

The method easily extends to two-dimensional tracking. For example, using two-dimensional orthogonal strip readout as charge collection electrode [13], the horizontal strips and the vertical strips are encoded multiplexing readout, respectively. By comparison of the conventional direct pixel readout [14] which can resolve multiple events in coincidence, the encoded multiplexing readout is limited to single events within the acquisition window which depends on the dead time of electronics. Unlike the direct pixel readout, the method can't handle the case of high incident flux,but it still is comparable with 10 kHz/cm$^2$ of the similar method [9]. However, the method has a great advantage that it can significantly reduce the number of readout channels. The method can readout $C_n^2$ strips by n readout channels while the conventional direct readout [15] needs $C_n^2$ channels. For a lage 50×50cm² GEM detector with 0.5mm pitch one-dimensional strips readout, the conventional direct readout needs 1000 readout channels, but this encoded method only needs about 50 readout channels.

A novel method of encoded multiplexing readout for micro-pattern gas detectors is presented in this work. This method is systematic and easily-extensible, offering a general way of encoding and decoding. The method is verified by a test of the 5×5 cm² THGEM detector equipped with the encoded anode readout PCB, where 166 strips are read out by 21 encoded readout channels. The test results show its general properties and performance under operation with 8 keV X-rays with 100 μm slit. It has a good linearity in its position reponse,and the spatial resolution RMS of the test system is about 260 μm.

Although it is based on the redundancy that at least two neighboring strips record the signal of a particle, with the robust detector such as MPGDs and low-noise electronics, it still has an attractive potential to build large area detectors and has a wide range of applications in and beyond particle physics. Moreover, it can also be used for the other detectors like MPGDs, Drift Chambers or scintillators.

*Acknowledgment*

The authors thank LIU Qian for his useful suggestions and discussions, and LUO Wen-Tai for his help during the tests.